\documentclass[desactivate]{aa}

\usepackage{graphicx}
\usepackage{txfonts}
\usepackage{hyperref}
\usepackage{natbib}
\usepackage{siunitx}
\usepackage{amsmath}
\usepackage[nolist,nohyperlinks]{acronym}
\usepackage{colortbl}
\usepackage{multirow}

\DeclareSIUnit\DIT{DIT}
\DeclareSIUnit\Jansky{Jy}
\DeclareSIUnit\mag{mag}
\DeclareSIUnit\mas{mas}

\acrodef{BCD}{beam commuting device}
\acrodef{OPD}{optical path difference}
\acrodef{TTP}{tip-tilt-piston}
\acrodef{VLTI}{Very Large Telescope Interferometer}
\acrodef{DL}{delay line}
\acrodef{DIT}{detector integration time}
\acrodef{AT}{Auxilliary Telescope}
\acrodef{UT}{Unit Telescope}
\acrodef{AO}{adaptive optics}
\acrodef{YSO}{young stellar object}
\acrodef{AGN}{active galactic nucleus}
\acrodefplural{AGN}{active galactic nuclei}
\acrodef{NGS}{natural guide star}
\acrodef{LGS}{laser guide star}
\acrodef{RMN}{reflective memory network}
\acrodef{FO}{feeding optic}
\acrodef{OB}{observing block}
\acrodef{STS}{star separator}
\acrodef{LBT}{Large Binocular Telescope}
\acrodef{DDL}{differential delay line}
\acrodef{BCDDL}{beam compressor differential delay line}
\acrodef{GPAO}{GRAVITY+ Adaptive Optics}

\definecolor{LightGray}{rgb}{0.8,0.8,0.8}

\makeatletter
\renewcommand*\aa@pageof{, page \thepage{} of \pageref*{LastPage}}
\makeatother

\begin{document}

   \title{GRAVITY for MATISSE}

   \subtitle{Improving the MATISSE performance with the GRAVITY fringe tracker}

   \author{
      J. Woillez \inst{1}\fnmsep\thanks{\email{jwoillez@eso.org}},
      R. Petrov \inst{2}\fnmsep\thanks{\email{romain.petrov@unice.fr}},
      R. Abuter \inst{1},
      F. Allouche \inst{2},
      P. Berio \inst{2},
      R. Dembet \inst{1},
      F. Eisenhauer \inst{3},
      R. Frahm \inst{1},\\
      F. Gonté \inst{1},
      X. Haubois \inst{4},
      M. Houllé \inst{2},
      W. Jaffe \inst{5},
      S. Lacour \inst{1},
      S. Lagarde \inst{2},
      J. Leftley \inst{2},\\
      B. Lopez \inst{2},
      A. Matter \inst{2},
      A. Meilland \inst{2},
      F. Millour \inst{2},
      M. Nowak \inst{6},
      C. Paladini \inst{4},\\
      T. Rivinius \inst{4},
      D. Salabert \inst{2},
      N. Schuhler \inst{1},
      J. Varga \inst{7},
      G. Zins \inst{4}
   }

   \authorrunning{Woillez et al.}

   \institute{
      ESO Headquarters, Karl-Schwarzschild-Str. 2, 85748 Garching bei München, Germany \and
      Université Côte d'Azur, Observatoire de la Côte d'Azur, CNRS, Laboratoire Lagrange, Bd. de l'Observatoire, CS 34229, 06304 Nice cedex 4, France \and
      Max-Planck-Institut für extraterrestrische Physik, Giessenbachstrasse, 85748 Garching bei München, Germany \and
      ESO Vitacura, Alonso de Córdova 3107, Vitacura, Casilla 19001, Santiago de Chile, Chile \and
      Leiden Observatory, Huygens Laboratory, Niels Bohrweg 2, 2333 CA Leiden, The Netherlands \and
      Kavli Institute for Cosmology, University of Cambridge, Madingley Road, Cambridge CB3 0HA, United Kingdom \and
      HUN-REN Research Centre for Astronomy and Earth Sciences, Konkoly Observatory, Konkoly-Thege Miklós út 15-17, H-1121 Budapest, Hungary
   }

   \date{Received ----; accepted ----}

  \abstract
   {MATISSE, the mid-infrared spectro-imaging instrument of VLTI, was designed to deliver its designed performance when paired with an external second-generation fringe tracker. Science observations started in 2019, demonstrating imaging capabilities and faint science target observations. Now, the GRAVITY fringe tracker stabilises the MATISSE fringes, which allows the use of all spectroscopic modes and improves sensitivity and data accuracy.}
   {We present how the MATISSE and GRAVITY instruments were adapted to make the GRAVITY fringe tracker work with MATISSE, under the umbrella of the aptly named GRA4MAT project, led by ESO in collaboration with the two instrument consortia.}
   {We detail the software modifications needed to implement an acquisition and observing sequence specific to GRA4MAT, including simultaneous fringe tracking and chopping and a narrow off-axis capability inspired by the Galactic Centre and exoplanet capability of GRAVITY. We explain the modified data collection and reduction processes. We show how we leveraged the recent fringe tracker upgrade to implement features specific to its use with MATISSE, for example mitigation of fringe jumps with an improved group delay control, and simultaneous fringe tracking and chopping with a new state machine.}
   {We successfully demonstrate significant improvements to the MATISSE instrument. Observations can now be performed at higher spectral resolutions of up to $R\sim3300$ and across the full LM bands at once. Long detector integration times, made possible with stabilised fringes, have improved the LM-band sensitivity by a factor of 10. Low flux biases in coherently reduced N-band data have been eliminated. The L-band transfer function is now higher and more stable. We finally illustrate the scientific potential of GRA4MAT with a preview of the first exoplanet observation made by MATISSE on \object{$\beta$ Pictoris b}.}
   {}


   \maketitle
%

\section{Introduction}
\label{Sec:Introduction}

A second-generation fringe tracker for the \ac{VLTI} was originally foreseen to support the second-generation instruments GRAVITY \citep{GRAVITY+2017} and MATISSE \citep{Lopez+2022}.
A call for studies for such an instrument was issued by ESO in late 2008.
In a context where available resources were being allocated in priority to the second-generation instruments, including the former PRIMA astrometry project \citep{Sahlmann+2013}, none of the proposed solutions \citep{Blind+2010,Meisner+2012} materialised.
Whereas GRAVITY came with an internal fringe tracker \citep{Menu+2012,Choquet+2014,Lacour+2019}, MATISSE came without the fringe tracker it needed to achieve its planned performance and its circumstellar disks and active galactic nuclei science objectives.
This is how the idea of using the GRAVITY fringe tracker to stabilise the optical path differences of MATISSE came to life, as an alternative to this second-generation fringe tracker.

GRAVITY for MATISSE, hereafter GRA4MAT, was first proposed by ESO in 2013 as an alternative to the second-generation fringe tracker.
The concept was supported by the MATISSE consortium and endorsed by the ESO council in 2014.
Subsequently, concept studies were prepared by ESO and the MATISSE consortium in 2014-2015, ahead of a 2015 phase A review, conducted under the umbrella of the \ac{VLTI} facility project \citep{Woillez+2015}.
As such, GRA4MAT is the last project of the second-generation \ac{VLTI}.
The project only acquired momentum in 2019, after the completion over the 2015-2016 period of the significant upgrade of the interferometer infrastructure \citep{Woillez+2015} required to host the GRAVITY and MATISSE instruments, and the installation in late 2018 of a new adaptive optics system \citep[][NAOMI]{Woillez+2019} on the auxiliary telescopes.
The project passed final design review in August 2019, following prototyping activities carried out shortly before the review.
GRA4MAT was implemented, tested, and commissioned in Paranal over the period September 2019--March 2023, including an extended pause due to the COVID-19 pandemic reaching Europe and Chile in 2020, delays resulting from the failure of the L-band dispersion wheel inside MATISSE between January 2021 and February 2022, and needed synergies with the GRAVITY fringe tracker real-time controller upgrade \citep{Abuter+2016,Nowak+2024} between April 2022 and March 2023.

GRA4MAT is not the first instrument to pair a near-infrared fringe tracker and a mid-infrared instrument.
On the Keck Interferometer \citep{Colavita+2013}, the two-telescope Keck Nuller \citep{Serabyn+2012} was the first instrument of this kind.
It was primarily used to survey exo-zodiacal lights around nearby bright stars \citep{MillanGabet+2011,Mennesson+2014}, and developed many of the techniques further extended by the GRA4MAT project.
On the \ac{LBT}, a similar nulling instrument (LBTI) was also developed \citep[][]{Defrere+2016} for a similar exo-zodiacal purpose \citep{Ertel+2020}.
On the \ac{VLTI}, one of the objectives of the PRIMA project \citep{Delplancke2008} was to use the FSU fringe tracker \citep{Sahlmann+2009} to support the two-telescope MIDI instrument \citep{Leinert+2003}, with a setup called MIDI+FSU \citep{Muller2012,Muller+2014}; however, it remained at the level of an experiment.

GRA4MAT can also be considered as a precursor to instruments currently under development.
The L-band nuller NOTT \citep[formerly Hi-5][]{Defrere+2018} of the Asgard visitor instrument suite proposal \citep{Martinod+2023} aims at bringing the nulling technique pioneered in the US to VLTI.
GRA4MAT even has connections with the mid-infrared ELT instrument METIS \citep{Brandl+2021}, where water vapour dispersion introduces a flat wavefront mismatch between wavefront sensing in the visible and high-contrast imaging in the thermal infrared.
As such, GRA4MAT provides direct measurements of this effect on baseline lengths comparable to the ELT diameter \citep{Absil+2022}.
This represents an interesting parallel between adaptive optics on a \SI{40}{\meter} diameter telescope and fringe tracking on a \SI{120}{\meter} baseline interferometer.

\section{Implementation}
\label{Sec:Implementation}

GRA4MAT is almost exclusively a software implementation; none of the hardware of GRAVITY, MATISSE, or \ac{VLTI} was modified.
However, it benefited from the upgrade of the GRAVITY fringe tracker, which also included a hardware component \citep{Abuter+2016,Nowak+2024}.

\subsection{Hardware implementation}

\begin{figure}
   \centering
   \includegraphics[width=0.9\linewidth]{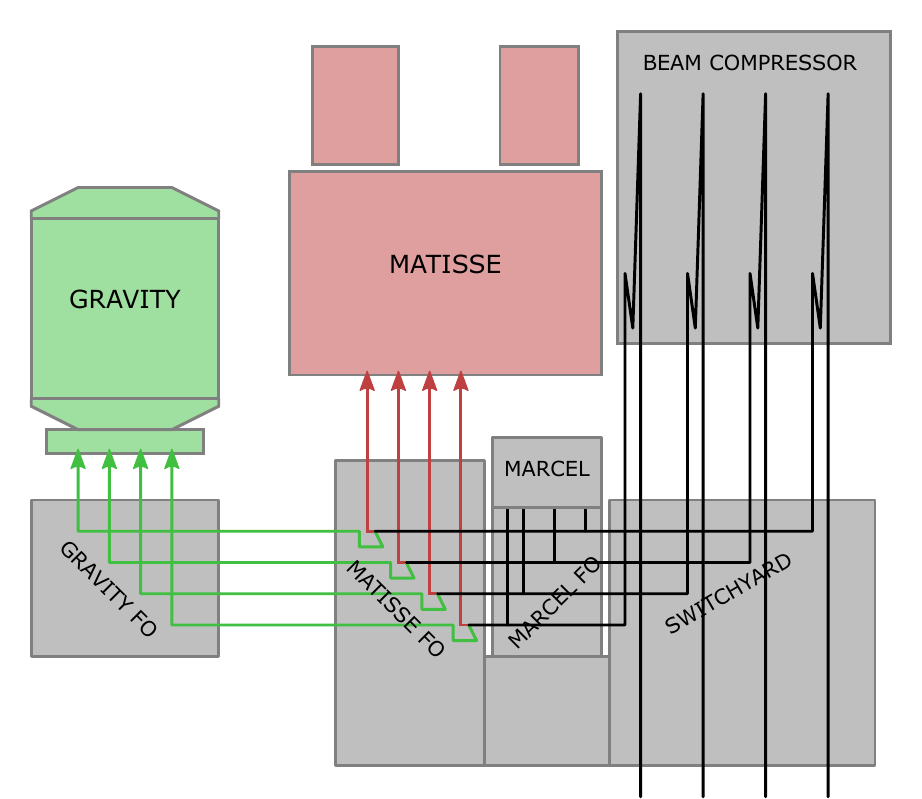}
   \caption{Layout of GRAVITY for MATISSE in the \ac{VLTI} laboratory. The infrared light beams reaching the laboratory are first reduced from \SI{80}{\milli\meter} to \SI{18}{\milli\meter} by the beam compressors, then sent by the switchyard towards the MATISSE \acp{FO}, and finally split by the \acp{FO} between the near-infrared reflected to GRAVITY (green) and the mid-infrared transmitted to MATISSE (red). The light split between MATISSE and GRAVITY is implemented with a dichroic in transmission and one mirror reflection towards MATISSE, and the same dichroic in reflection and three mirror reflections towards GRAVITY. MARCEL is a coherent calibration source that can be used to co-phase MATISSE and GRAVITY.}
   \label{Fig:Layout}
\end{figure}

The layout of GRA4MAT is the layout of MATISSE and GRAVITY in the \ac{VLTI} laboratory, illustrated in Fig. \ref{Fig:Layout}.
The GRA4MAT set-up uses the same MATISSE feeding optics that split the LMN bands for MATISSE and the K band for IRIS, the infrared tip-tilt sensor of \ac{VLTI}.
This time, the KH bands are sent to GRAVITY where the K band is used for fringe tracking and the H band for field guiding.
The simple long-wavelength-reflecting and short-wavelength-transmissive dichroic plate, dictated by the relative position of the two instruments, could not be manufactured for technical feasibility reasons.
Instead, fold mirrors are used to operate MATISSE in transmission and GRAVITY or IRIS \citep{Gitton+2004} in reflection.
The dichroic and fold mirrors are inserted in the beam only when MATISSE is operating in standalone or GRA4MAT modes; to switch from MATISSE standalone to GRA4MAT, the GRAVITY \acp{FO} are inserted in the beam.

\subsection{Software implementation}

The software architecture of GRA4MAT considers GRAVITY as a subsystem of MATISSE.
When the MATISSE instrument operates in GRA4MAT mode, its state also depends on the state of GRAVITY.
The MATISSE acquisition and observation sequencers directly control GRAVITY without going through the Interferometer Supervisory Software, the interface to the \ac{VLTI} infrastructure for all \ac{VLTI} instruments.
The functions available from the GRAVITY instrument and fringe tracker workstations are accessed directly by MATISSE, whereas functions only available from GRAVITY local control units are accessed through a gateway running on the GRAVITY workstation.
Feedback from GRAVITY is obtained through live database links from the GRAVITY workstation to the MATISSE workstation.

\subsection{Acquisition sequence}
\label{SSec:AcquisitionSequence}

The acquisition sequence starts with MATISSE sending the target coordinates to the \ac{VLTI} and its telescopes.
Once the visible adaptive optics loop closes on the guide star, the infrared light reaches the \ac{VLTI} laboratory and the target acquisition by GRAVITY can start.

Rather than re-implementing the GRAVITY acquisition procedure in the MATISSE instrument, which would require access from MATISSE to many low-level GRAVITY functions, MATISSE prepares an acquisition \ac{OB} following the single off-axis template and remotely triggers its execution on GRAVITY using the tool called Bob\_Stephanie.\footnote{`Bob' is the Broker of Observations Block, well known to VLT users. Usually, one single Bob runs per instrument. Since another Bob runs on the GRAVITY side, in the GRA4MAT case we gave it another name: Stephanie.}
This \ac{OB} skips sending target coordinates to an already tracking \ac{VLTI}, but configures GRAVITY for single off-axis operations to send all the K-band light to the fringe tracker.
This represents a two-fold transmission gain over the on-axis mode usually employed by GRAVITY for single-target observations. This choice optimises the fringe tracker limiting magnitude, but prevents any light from reaching the GRAVITY spectrometer.

The acquisition sequence for GRAVITY follows the standard procedure.
The acquisition camera field loop adjusts the internal tip-tilt mirror of the instrument to couple the target light into the single-mode fibres of the fringe tracker.
A small synchronous modulation of the same internal mirror, and demodulation of the coupled flux, removes any pointing mismatch between the single-mode fibres and the acquisition camera.
Searching and tracking fringes completes the GRAVITY acquisition procedure.
Then the acquisition on MATISSE resumes.

First, MATISSE switches the actuator used by GRAVITY for field control from the internal \ac{TTP} actuator to the common-path field steering mirrors of the telescope \aclp{STS} \citep[\acsu{STS},][]{Nijenhuis+2005}.
This makes GRAVITY stabilise the field for both instruments.
MATISSE first offsets the internal tip-tilt actuator of GRAVITY to a reference value that corresponds to the co-alignment between the two instruments.
This offset is compensated by the common-path field steering mirror of the \ac{STS} to keep the target position unchanged on the fringe tracker fibres.
Then MATISSE performs a field acquisition measurement and sends combined offsets to the internal tip-tilt mirrors of GRAVITY and the \acp{STS} to refine the co-alignment between the two instruments.
After this operation, since the GRAVITY field derotator remains static in single off-axis mode, the two instruments remain co-aligned all through the night, and this acquisition does not have to be repeated after the first object of the night.

Second, MATISSE switches the actuator used by GRAVITY for piston control from the internal tip-tilt to the common-path \ac{VLTI} delay lines.
This switch comes with an adjustment of the actuator model used by the predictive Kalman controller of the fringe tracker.
The \ac{VLTI} delay lines have a \SI{500}{\micro\second} longer latency (see Fig. \ref{Fig:Latency}).
Again, this makes GRAVITY stabilise the piston for both instruments.
MATISSE completes the acquisition sequence with an offset to its internal delay lines to match the optical path differences of GRAVITY.

\subsection{Coherent flux observations}
\label{SSec:CohObs}

The GRA4MAT observation sequence remains identical to that with MATISSE standalone.
After a sky measurement, non-chopped fringes are recorded for the requested number of \ac{BCD} cycles \citep[see section 2.4 in][]{Lopez+2022}.

Except for the piston actuator replaced by the main delay lines, the fringe tracker operates for MATISSE the same way it does for GRAVITY.
The only critical aspect is related to how GRA4MAT configures the fringe tracker for the management of dispersion and the mitigation of fringe jumps.
Since its upgrade \citep{Nowak+2024}, the fringe tracker is capable of periodically updating the set point of its group delay control loop.
A regular update of this set point is critical to follow the slowly evolving dispersion, and keeping fringe jumps under control.
GRA4MAT uses this feature to schedule the updates between the MATISSE \acp{DIT}, the same way GRAVITY standalone does it between its own \acp{DIT}.
When the L-band \acp{DIT} are shorter than \SI{1}{\second}, GRA4MAT schedules the updates between the OPD modulation cycles applied by MATISSE on the L-band side of the instrument.
When the L-band \acp{DIT} are longer than \SI{1}{\second}, the updates are scheduled before each L-band \ac{DIT}.
This is done because the group delay set point is constant and is not able to follow the evolving dispersion over hundreds of seconds; its validity in time is therefore limited.
For long L-band \acp{DIT}, the fringe tracker therefore does not keep the phase stable for the duration of a MATISSE \ac{OPD} cycle, and coherent integrations by the MATISSE pipeline must consider the group delay tracked by GRAVITY.

\subsection{Narrow off-axis offsets}
\label{Sec:NarrowOffAxisOffsets}

The narrow off-axis mode of GRA4MAT was not part of the initial objectives of the project.
It was added along the way in light of the pioneering Galactic Centre results \citep[e.g.][]{GRAVITY+2018} and exoplanet results \citep[e.g.][]{GRAVITY+2019} obtained with GRAVITY.
With this mode, MATISSE is now able to observe any faint target within a \ac{VLTI} field of view (\SI{2}{\arcsecond} and \SI{4}{\arcsecond} diameter with \acp{UT} and \acp{AT}, respectively) centred on a bright fringe tracking target.

To perform a narrow off-axis observation, the user provides in the observing template a sequence of cumulative on-sky offsets starting from the fringe tracker target.
This allows multiple possibly fainter targets to be sequentially observed, around the fringe tracking target serving as a calibrator.
Such a sequence is illustrated in Table \ref{Tab:NarrowOffAxisOffsets}.
The requested relative offsets are applied in tip-tilt and piston, by the MATISSE observation sequencer, at the start of each \ac{BCD} cycle.

\begin{table}
   \fontsize{8}{9.5}\selectfont
   \centering
   \caption{Typical narrow off-axis offsetting sequence for a planet observation.}
   \label{Tab:NarrowOffAxisOffsets}
   \begin{tabular}{|cc|cc|c|}
      \hline
              \multicolumn{2}{|c|}{Relative offset}       &        \multicolumn{2}{c|}{Exposure offset}        &             \\
              \multicolumn{2}{|c|}{(from template)}       &        \multicolumn{2}{c|}{(in fits header)}       &             \\
       RA [\si{\milli\arcsec}] & DEC [\si{\milli\arcsec}] & RA [\si{\milli\arcsec}] & DEC [\si{\milli\arcsec}] &    Target   \\
      \hline
      \hline
               $  0 $          &          $  0 $          &         $  0 $          &          $  0 $          &     Star    \\
               $+284$          &          $+462$          &         $+284$          &          $+462$          &    Planet   \\
               $-568$          &          $-924$          &         $-284$          &          $-462$          & Anti-planet \\
               $+568$          &          $+924$          &         $+284$          &          $+462$          &    Planet   \\
      \hline
   \end{tabular}
   \tablefoot{The planet is located at (\SI{284}{\mas}, \SI{+462}{\mas}) offset from its host star. The relative offsets in RA and DEC, specified in the observation template, have as many entries as requested exposure cycles, four in the presented sequence. The cumulated offset is reported in FITS header under keywords \textsc{SEQ.OFFSET.ALPHA} and \textsc{SEQ.OFFSET.DELTA}. This sequence corresponds to the publicly available commissioning dataset on \object{$\beta$ Pictoris b} (see \url{http://www.eso.org/sci/activities/instcomm/matisse.html}, Run ID: 60.A-9257(H)). The intent of the offset sequence was to observe the planet, calibrate it with the star, and investigate off-axis stellar leakage at the anti-planet position.}
\end{table}

In tip-tilt, this offset is a combined offset between the common-path field steering mirrors of the \acp{STS} and the GRAVITY-only internal tip-tilt mirrors.
As a result, the MATISSE pinhole moves to the requested offset while the GRAVITY fringe tracker remains in place.
It should be noted that this offset strategy does not continuously compensate for field rotation.
Although this is not an issue with the \acp{AT} that deliver a stable field orientation to the interferometric laboratory, this affects the autonomy for the \acp{UT} and their rotating fields.
Therefore, a recalculation of the narrow off-axis offset is performed before each \ac{BCD} state, leaving the \acp{UT} field rotation uncompensated over no more than \SI{\sim 1}{\minute} exposure intervals.

In piston, the \ac{OPD} offset is performed with a combined on-sky offset between the common-path \ac{VLTI} delay lines and the GRAVITY-internal fibred delay lines.
An offset of twice the requested field offset is added to the fibred delay lines trajectory.
The factor is explained by the GRAVITY fibred delay lines operating with half the offset on the fringe tracker side and half on the science camera.
As a result, the fringe tracker fibred delay lines follow the opposite of the field offset.
To preserve the fringe position on GRAVITY, the main delay lines must compensate for this offset.
This is done by updating the fringe tracker fringe offsets sent to the main delay lines.
Under the control of the fringe tracker, the main delay lines end up following an offset trajectory that corresponds to the exposure offset, without any autonomy limitation.

\subsection{Visibility observations}
\label{SSec:VisObs}

An absolute flux measurement is required to convert coherent flux measurements to visibilities.
Owing to the stronger thermal background in the mid-infrared, this measurement tends to be more difficult to obtain accurately; this has been a challenge for all thermal interferometers.
The Keck Interferometer Nuller \citep{Serabyn+2012} could not rely on chopping secondary mirrors on both Keck telescopes, and had to implement an interferometric chopping strategy on the short baselines corresponding to the two halves of a same telescope.
The MIDI+FSU experiment \citep{Muller+2014} focused on faint coherent flux measurements and did not attempt to measure chopped photometry.
LBTI, where PhaseCam \citep{Defrere+2014} is the HK-band fringe tracker and NOMIC \citep{Hoffmann+2014} and LMIRcam \citep{Skrutskie+2010} are the thermal science cameras, is probably the facility closest to GRA4MAT.

The initial GRA4MAT chopping implementation, offered for general use with the \acp{AT} starting in October 2021 (P108), relied on the GRAVITY fringe tracker capability to react to flux losses on all four telescopes and transition to a so-called SKY state, where \ac{OPD} control would become frozen.
This reactive chopping was not completely satisfactory, due to a poor control of the transitions between the sky and target states.
While chopping, the fringe tracker would easily fall out of lock and become unable to find fringes again.
Already present on the \acp{AT}, this situation was more severe on the \acp{UT}, to a point where chopping could not be offered with the large telescopes.
A better implementation was needed.

The situation could only be improved after the upgrade of the fringe tracker real-time controller \citep{Abuter+2016,Nowak+2024} in November 2022.
In January 2023 an active chopping state was integrated into the new real-time software.
The chopping of the fringe tracker behaves like any other chopping element of \ac{VLTI}.
Its sequence is fully specified by an absolute start time, a period, and a sky-to-target ratio.
When chopping is active, the timing of the group delay control loop set point is adjusted to fall right at the end of each TARGET phase, just before the transition to SKY.
This was chosen because this is when the fringe tracker has the highest probability of having a correct estimation of the dispersion between phase delay and group delay.
Updating the set point at the start of the TARGET phase would have required to first wait for the dispersion estimators to build S/N, increasing further the time needed to recover proper fringe tracking performance.
For this scheduling strategy to work, the dispersion must remain within $\pm\lambda/2$ for the duration of a full chopping cycle.

Although the target-to-sky transition happens at the same time for all chopping participants, the sky-to-target transition must happen in sequence.
First the chopping mirror (deformable mirror on \acp{AT} and telescope secondary mirror on \acp{UT}) must move back to the target position, then the adaptive optics must close the loop and take a few iterations to deliver the correction level needed for the fringe tracker to resume control.
For the current implementation, this delay is set to \SI{15}{\milli\second} after measuring the time it takes for the flux injected into the fringe tracker to return to its typical level.
The fringe tracker finally needs some additional time to reacquire the fringe identified before the transition to sky.
In total, we implemented a +\SI{250}{\milli\second} delay from the start of the target phase to when the L band starts its integration.
The majority of the \SI{250}{\milli\second} delay is consumed by the fringe tracker slewing back to the reference fringe.
This constraint is set by the length of the fringe tracker group delay averaging window of 150 samples or \SI{165}{\milli\second} at a detector period of \SI{1.1}{\milli\second} and by the speed limit of the group delay corrections.
The maximum chopping frequency is set to <\SI{0.5}{\hertz} (>\SI{1}{\second} on sky and >\SI{1}{\second} on target) and practically limited by the fringe recovery time.
The fixed +\SI{250}{\milli\second} delay has been tuned for the \SI{1.1}{\milli\second} period of the fringe tracker.
As such, chopping with GRA4MAT is not supported when the fringe tracker operates more slowly, and the following limits must be respected: K < \SI{7}{\mag} with the \acp{AT}, and K < \SI{10}{\mag} with the \acp{UT}.

Chopping with GRA4MAT is offered on the \acp{AT}, but not yet on the \acp{UT}.
At the time of writing, the limitation on the \acp{UT} results from the lower performance of the Adaptive Optics system.
Despite the active chopping improvements, the fringe tracker tends to lose fringes more often on the \acp{UT}.
This situation is expected to improve when the forthcoming \ac{UT} adaptive optics upgrade is deployed \citep{GRAVITY++2022B}.

\subsection{Data collection}
\label{SSec:DataCollection}

The GRAVITY fringe tracker supports the MATISSE performance at two levels.
First, it actively stabilises the optical path difference and improves the MATISSE coherence beyond the atmospheric limits up to the background limited integration times of MATISSE.
Second, it contributes chromatic complex coherence and flux in the K band to the data recorded by MATISSE.
This information is generated directly by the fringe tracker real-time controller, published through the \ac{RMN}, and recorded by the \ac{RMN} recorder \citep{Abuter+2008} under the control of the MATISSE parallel exposures.
These records are part of the dataset sent to the ESO archive and used by the instrument pipeline for coherent integrations and frame selection.

MATISSE does not record the raw pixels from the GRAVITY fringe tracker detector, but only the COHERENCE quantity\footnote{The fringe tracker P2VM output is available at each fringe tracker \ac{DIT} in the COHERENCE column of the IMAGING\_DATA\_FT extension of the DPR.TYPE=*,RMNREC FITS files.} at the output of the P2VM.
Specifying the fast axis first, the dimensionality of COHERENCE is (real part for 6 baselines coherence + imaginary part for 6 baselines coherence + 4 telescopes fluxes) * 4 wavelengths = 64 values.
Only the four brightest central wavelengths are recorded (out of the six read from the detector) with a wavelength sampling [\qtylist[list-units = single]{2.076; 2.170; 2.270; 2.374}{\micro\meter}].
This approach is the closest way of obtaining scientifically useful data from the fringe tracker telemetry while remaining compatible with the ESO dataflow constraints.
However, the P2VM matrix of the fringe tracker is not regularly updated and the usability for science is not validated.
Recording the raw pixels would have required dealing with calibrations of another instrument inside the MATISSE dataflow and pipeline, which was beyond the scope of the project.

\subsection{Data reduction}
\label{SSec:DataReduction}

All data reductions specific to GRA4MAT are implemented in the same MATISSE pipeline that reduces MATISSE standalone data \citep[see][section 3.3]{Lopez+2022}.
We present here only the additions to support the GRA4MAT mode.
The GRA4MAT-specific behaviour is activated when the fringe tracker telemetry is provided to the pipeline.

The pipeline first uses the GRAVITY telemetry to detect fringe jumps (i.e. when the fringe tracker group delay moves by more than half a wavelength).
The recorded MATISSE frames that contain jumps are flagged with TARTYP=J, and dismissed from the remaining computations of the pipeline.
The extent of the flagging depends on the MATISSE spectral bands.
In N band, all the frames of a complete ten-step modulation cycle are rejected, corresponding to a period of \SI{260}{\milli\second} in low resolution and \SI{840}{\milli\second} in high.
In L band, only the frame corresponding to the detected jump is rejected.
We note that even if fringe jumps are detected in the group delay of one baseline, all baselines are affected by the flag.

\begin{figure}
   \centering
   \includegraphics[width=\linewidth]{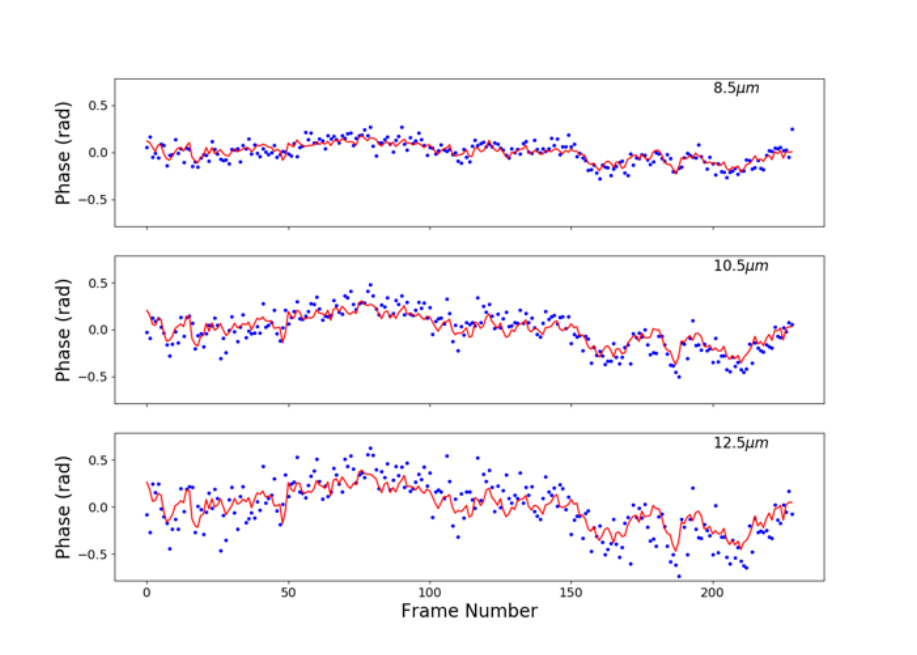}
   \caption{Phase estimated directly from MATISSE data (blue dots) and predicted from the fringe tracker data (red line) on target 7 Cet (38 Jy) observed with baseline D0-C1 at three different wavelengths across the N band (\SI{8.5}{\micro\meter}, \SI{10.5}{\micro\meter,} and \SI{12.5}{\micro\meter}).}
   \label{Fig:NBandPhasePrediction}
\end{figure}

In the N band, the thermal background contribution dominates many astrophysical sources.
The N-band detector integration time is then set to avoid any thermal background saturation: \SI{20}{\milli\second} in low spectral resolution, \SI{75}{\milli\second} in high.
As such, coherent integration in the pipeline are necessary to improve the S/N of interferometric observables.
In the GRA4MAT mode, the estimation of the N-band phase is derived from the fringe tracker K-band measurements.
Even if the fringes are stabilised by the fringe tracker, the N-band \ac{OPD} fluctuates at a level of a few microns, due to the dispersive water vapour turbulence.
The MATISSE pipeline follows the procedure of \citet{Koresko+2006} to extrapolate the K-band measurements to the N band: the differential column density of water vapour is estimated from the K-band phase delay and group delay measurements, and linearly converted into a prediction of the phase delay and group delay in the N band.
We recall the equations involved in this conversion:
\begin{align}
   PD_\lambda &= PD_K + \left[ N^w_\lambda - N^w_K \right] \gamma^w_K D_K \label{Eq:1},\\
   D_\lambda  &= \frac{\gamma^w_K }{ \gamma^w_\lambda} D_K\label{Eq:2}.
\end{align}
Here the phase delay $PD$, the group delay $GD$, and the dispersion $D=PD-GD$ are in units of optical length; $N^w$ is the water vapour refractivity; and $\frac{1}{\gamma^w} \equiv \lambda \frac{\partial N^w}{\partial\lambda}$ is its group equivalent.
Table \ref{Tab:CoPhasingParam} gives numerical values for the above parameters, derived from the refractive indexes of \citet{Colavita+2004}.
Figure \ref{Fig:NBandPhasePrediction} illustrates the K-to-N prediction process on a target bright enough in N band to estimate the phase directly from the MATISSE data.

A similar K-to-L correction is currently not implemented in the MATISSE pipeline.
Since the water vapour decorrelation from K band to L band is not as strong as to N band \citep[Table 5 in][]{Colavita+2004}, such a correction was not deemed critical to the project.
However, recent developments have shown that a residual phase jitter correction would certainly help stabilise the MATISSE transfer function, and open the door to higher contrast observations.

\begin{table*}
   \centering
   \caption{\centering Water vapour co-phasing parameters.}
   \label{Tab:CoPhasingParam}
   \begin{tabular}{|c|c|c|c|c|}
      \hline
      Band & $N^w$ & $\frac{1}{\gamma^w} \equiv \lambda \frac{\partial N^w}{\partial\lambda}$ & $\left[N_\lambda^w-N_K^w \right] \gamma_K^w$ & $\frac{\gamma_K^w}{\gamma_\lambda^w}$ \\
      \hline
      \hline
      K (\SI{ 2.2 }{\micro\meter}) & \num{90.29e-8} & \num{ -9.13e-8} & 0.000 & 1.000 \\
      L (\SI{ 3.75}{\micro\meter}) & \num{87.95e-8} & \num{-13.05e-8} & 0.256 & 1.429 \\
      M (\SI{ 4.75}{\micro\meter}) & \num{83.49e-8} & \num{-30.24e-8} & 0.745 & 3.312 \\
      N (\SI{10.5 }{\micro\meter}) & \num{64.91e-8} & \num{-71.43e-8} & 2.780 & 7.824 \\
      \hline
   \end{tabular}
   \tablefoot{The numbers above are derived from the refractivity indexes calculated by \citet{Colavita+2004}. They are meant to be used with Eq.~\ref{Eq:1} and \ref{Eq:2}, following the strategy of \citet{Koresko+2006}.}
\end{table*}

\section{Fringe tracking in the GRA4MAT context}
\label{Sec:FringeTracking}

\begin{figure}
   \centering
   \includegraphics[width=\linewidth]{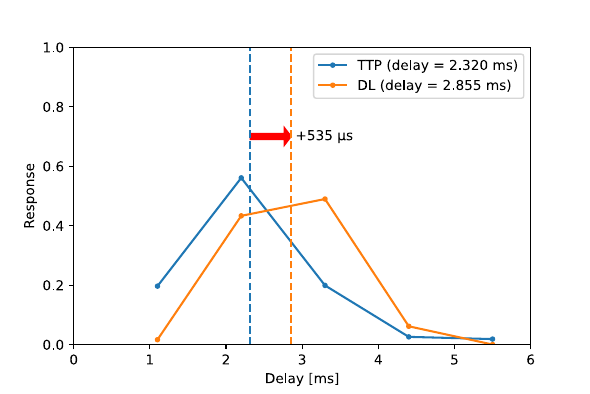}
   \caption{Comparison of the normalised impulse response (solid lines) at \SI{1.1}{\milli\second} integration period between \ac{TTP} used by GRAVITY standalone, and the main \ac{VLTI} \ac{DL} used by GRA4MAT. The weighted average of the impulse response (vertical dashed lines) show an additional latency of \SI{\sim500}{\micro\second}. The responses were measured in 2023 after the fringe tracker real-time controller upgrade.}
   \label{Fig:Latency}
\end{figure}

In this section we present aspects of the GRAVITY fringe tracker performance, when supporting MATISSE, an instrument operating at a different wavelength.
The GRAVITY fringe tracker uses the main \ac{VLTI} delay lines as \ac{OPD} actuators when operating with GRA4MAT.
The delay lines are known to be affected by extra latency, reducing the \ac{OPD} control performance when compared with the internal \ac{TTP} actuator of GRAVITY.
Figure \ref{Fig:Latency} compares the responses of the two actuators and shows a \SI{\sim500}{\micro\second} additional latency.
This actuator response, a necessary parameter of the controller implementation \citep[see][]{Nowak+2024}, is updated each time the fringe tracker switches between GRAVITY and GRA4MAT.
The consequences for MATISSE are limited, however, since it operates at longer wavelength, and this small fringe tracking performance degradation is by far not the dominant factor.

As of 2023, GRA4MAT operates with NAOMI \citep{Woillez+2019} on the \acp{AT} and MACAO \citep{Arsenault+2004} on the \acp{UT}.
We recall that NAOMI has significantly smaller sub-apertures than MACAO (\SI{\sim45}{\centi\meter} versus \SI{\sim1}{\meter}).
In the high flux regime, the Strehl delivered by NAOMI is therefore higher than that delivered by MACAO.
In addition, due to a reduced level of turbulence aliasing in the sub-apertures of the wavefront sensor, the NAOMI performance does not degrade as quickly as MACAO when the seeing conditions become poorer than median.
Because of this, we obtain very different fringe tracking behaviour when comparing \acp{AT} and \acp{UT}.
On the \acp{AT}, fed with a higher and more stable Strehl, the fringe tracker can continuously track the same fringe without jumping, 
whereas on the \acp{UT} the lower and more variable Strehl results in flux dropouts on the fringe tracker, which significantly increase the probability of fringe jumps.
The connection between median Strehl, flux losses, and fringe jumps is best illustrated in \citet{Woillez+2019}.

\begin{figure}
   \centering
   \includegraphics[width=\linewidth]{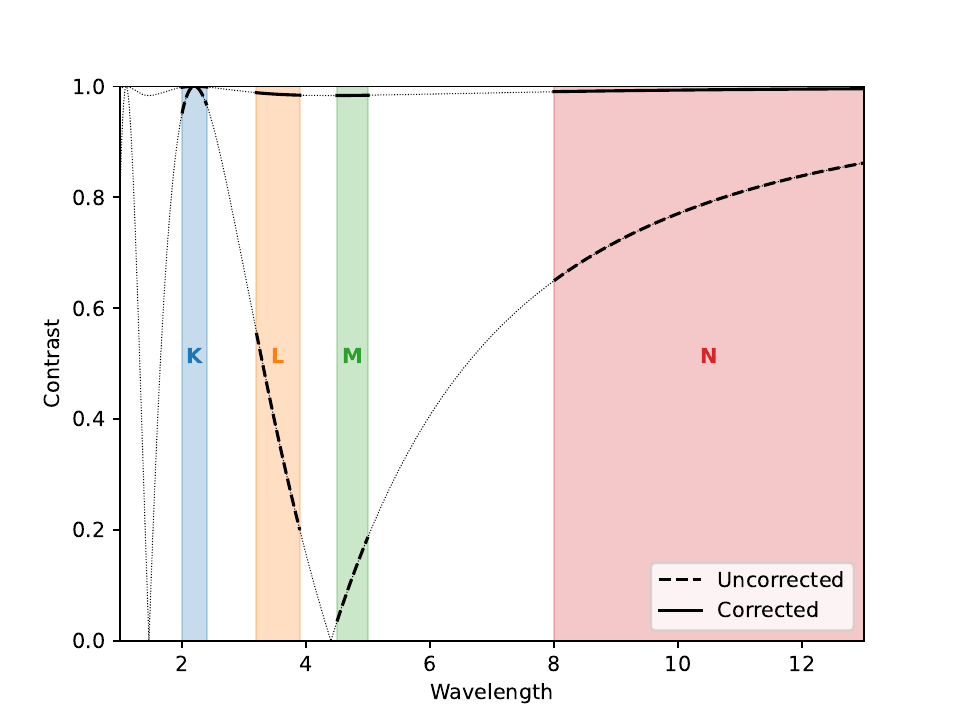}
   \caption{Comparison of the \SI{10}{\second} fringe contrast, obtained across the KLMN bands, between before the group delay controller upgrade when fringe jumps were left \textbf{uncorrected} and after the upgrade when \textbf{corrected} in \SI{82.5}{\milli\second}. In the uncorrected case, the K-band fringe jump is simulated in the middle of the \ac{DIT}, the worst case scenario. The K-band fringe jumps correction brings marginal benefits on the sides of the K band, but makes a very significant difference in the LMN bands of MATISSE.}
   \label{Fig:JumpLoss}
\end{figure}

The fringe jumps were not as bad for GRAVITY standalone as they are for GRA4MAT.
GRAVITY operates a spectrometer in the same band as its fringe tracker, and a fringe jump only has limited effects on the sides of the K band.
As shown in Fig. \ref{Fig:JumpLoss}, the situation is very different for GRA4MAT where a fringe jump in K band generates a fringe phase shift very different from \SI{360}{\degree}: \SI{230}{\degree} at \SI{3.5}{\micro\meter} (L band), \SI{160}{degree} at \SI{5}{\micro\meter} (M band), and \SI{80}{\degree} at \SI{10}{\micro\meter} (N band).
In addition, since the fringe jumps can happen anywhere inside a \ac{DIT}, the resulting fringe contrast becomes highly variable and impossible to calibrate without knowing when these jumps happen.

Initially, the GRAVITY fringe tracker did not have any mechanism in place to mitigate the impact of the fringe jumps.
As described in \citet{Lacour+2019}, the group delay controller had a dead zone of $\pm 3 \pi$ (or $\pm1.5\lambda$), leaving room for three stable tracking solutions that fringe jumps could randomly explore.
In November 2022, the real-time controller was upgraded, replacing the slower VME-based implementation \citep{Lacour+2019} by a faster Linux workstation \citep{Abuter+2016,Nowak+2024}.
This faster fringe tracker made possible the deployment of an improved group delay control algorithm that reduces the dead zone to less than a wavelength, leaving room for exactly one stable tracking solution for the duration of a GRAVITY spectrometer or MATISSE L-band \ac{DIT}.
The constant set point of the group delay controller is now updated between the MATISSE \acp{DIT}.
The time between updates must remain below the evolution timescale of the dispersion caused by dry air and wet air turbulence.
If not, when the dispersion shifts by more than a wavelength, the fringe tracker jumps without correction.

With this new implementation, the fringe jumps happen as often as before, but are detected and corrected in less than half the group delay estimator filter length of 150 elements (i.e. \SI{82.5}{\milli\second} when the integration period is the standard \SI{1.1}{\milli\second}).
Figure \ref{Fig:JumpLoss} illustrates the theoretical improvement of the fringe contrast, and fringe contrast stability, achieved with the new implementation.

The amount of water vapour turbulence varies with time and makes the maximum L-band \ac{DIT} dependent on the atmospheric conditions.
However, we do not yet have any analysis comparing the dispersion fluctuations to the precipitable water vapour or humidity, which would let the observer identify when conditions are compatible with absolutely no dispersion-induced fringe jumps.
Preliminary observations indicate that a line-of-sight precipitable water vapour below \SI{5}{\milli\meter} could be an acceptable threshold.

We conclude this section by recalling that the K-band limiting magnitudes of GRA4MAT are identical to GRAVITY operating in one of the off-axis modes.
The expected three-magnitude difference between \acp{AT} and \acp{UT} remains unobserved for now. This is a consequence of the lower relative performance of the \ac{UT} adaptive optics.

\section{MATISSE performance improvements}
\label{Sec:MatissePerformance}

In this section, we present aspects of the MATISSE performance improvements obtained when operating with the GRAVITY fringe tracker.
A comparison between MATISSE standalone and with GRA4MAT is also presented in Table \ref{Tab:PerformanceSummary}.

\subsection{Full LM-band spectral coverage}
\label{SSec:ImprovedCoverage}

\begin{figure*}
   \centering
   \includegraphics[width=\linewidth]{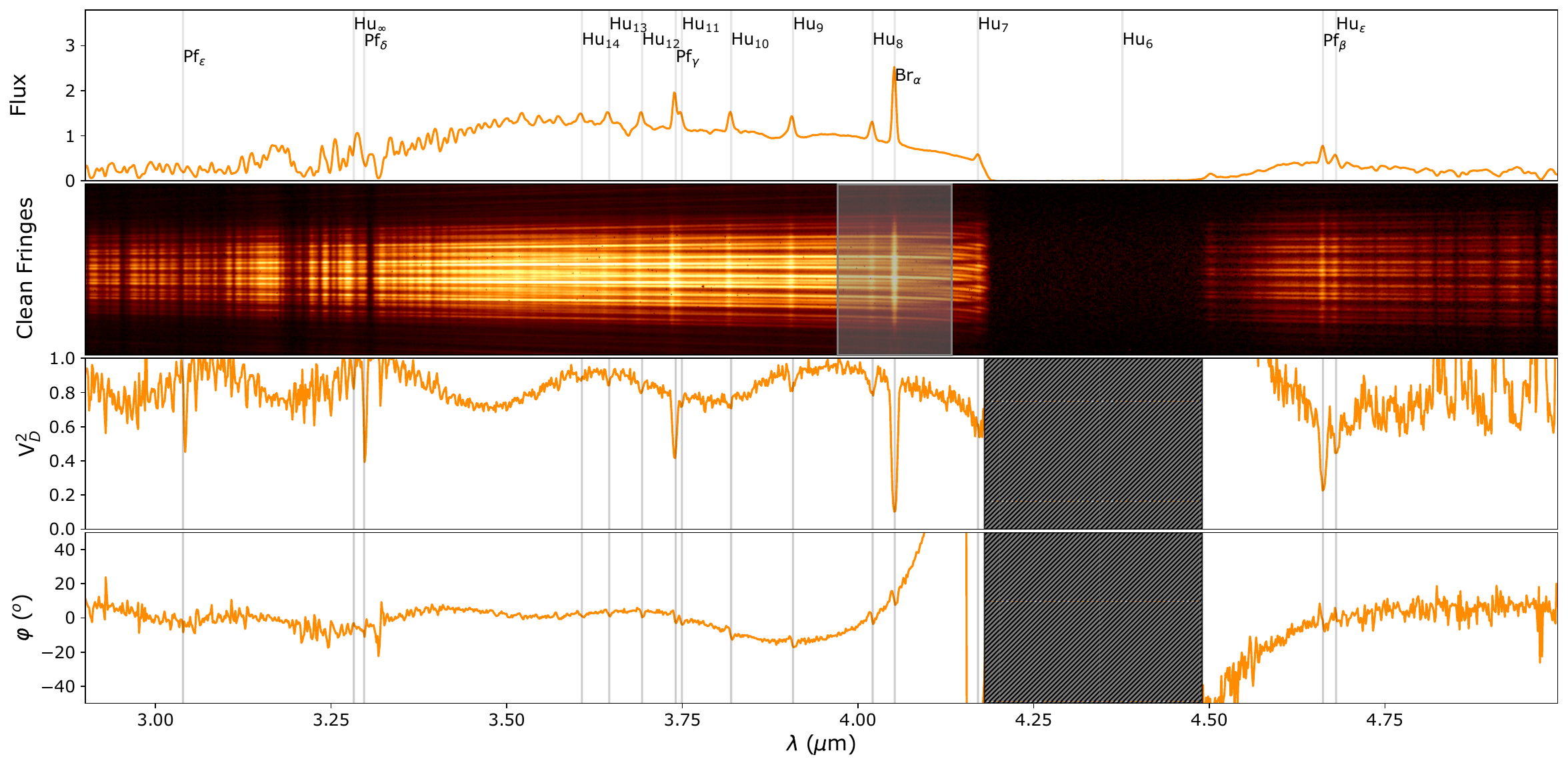}
   \caption{Medium-resolution LM-band observations of the Be binary $\delta$ Cen with GRA4MAT. The figure, already presented in \citet{Lopez+2022}, illustrates the improved instantaneous spectral coverage when operating with the fringe tracker. The long GRA4MAT-enabled \acp{DIT} allow the full LM bands to be captured, whereas MATISSE standalone is limited to a width of \SI{0.16}{\micro\meter}, corresponding to the grey outline for a Br$\alpha$ central wavelength. Top to bottom: Chromatic flux, clean fringes as observed on the LM-band detector, chromatic squared visibility, chromatic phase.}
   \label{Fig:DeltaCen}
\end{figure*}

MATISSE employs a HAWAII-2RG detector for the LM-band observations.
Its readout speed is limited by a slow pixel clock of \SI{100}{\kilo\hertz}.
This corresponds to a full detector readout time of \SI{1.3}{\second}, well beyond the atmosphere coherence time.
Windowing the detector is the only solution to reduce the readout time, at the expense of spectral resolution and/or wavelength coverage.
At an integration time of \SI{0.111}{\second}, within the L-band atmospheric coherence time, a full LM-band spectral coverage can only be achieved at low spectral resolution.
For higher spectral resolutions (medium and high), only a small fraction of the wavelength coverage can be obtained.
The highest spectral resolution (high+) is not even offered with MATISSE standalone.

The situation changes completely with GRA4MAT.
When fringes are stabilised, the LM-band detector integration time can be increased to \SI{1.3}{\second} without coherence loss and the full spectrum can be read.
Illustrated in Figure \ref{Fig:DeltaCen}, the observation of the spectroscopically rich binary Be star $\delta$ Cen shows this improved wavelength coverage.
To obtain the same spectral coverage without a fringe tracker, one would need a sequence of about 11 consecutive observations.
This represents a ten-fold gain in observing efficiency.

\subsection{Improved LM-band sensitivity}
\label{SSec:ImprovedLMSensitivity}

When the MATISSE fringes are stabilised by the GRAVITY fringe tracker, one would ideally want to increase the \acp{DIT} to a regime where the detector readout noise becomes negligible with respect to the thermal background noise.
However, to keep the probability of fringe jumps to a minimum within each \ac{DIT}, and to preserve the OPD modulation capability within an exposure, having the thermal background noise slightly above the detector readout noise is sufficient.
Table \ref{Tab:BackgroundReadoutNoiseDIT} gives the \acp{DIT} at various wavelengths that correspond to the background noise matching the readout noise.

\begin{table}
   \centering
   \caption{MATISSE L-band \acp{DIT} where the thermal background noise matches the readout noise.}
   \label{Tab:BackgroundReadoutNoiseDIT}
   \begin{tabular}{|c|c|c|c|c|}
      \hline
      L band                  &       low          &       medium       &       high         &       high+        \\
      \hline
      \hline
      \SI{3.0}{\micro\meter}  & \SI{0.08}{\second} & \SI{3.60}{\second} & \SI{14.0}{\second} & \SI{232}{\second}  \\
      \SI{3.5}{\micro\meter}  & \SI{0.02}{\second} & \SI{0.90}{\second} & \SI{3.50}{\second} & \SI{58}{\second}   \\
      \SI{4.05}{\micro\meter} & \SI{0.01}{\second} & \SI{0.45}{\second} & \SI{1.75}{\second} & \SI{29}{\second}   \\
      \hline
      \hline
      GRA4MAT                 &  \SI{1}{\second}   &  \SI{10}{\second}  &  \SI{10}{\second}  &  \SI{10}{\second}  \\
      \hline
   \end{tabular}
   \tablefoot{The \acp{DIT} are taken from \citep{Lopez+2022} and given as a function of wavelength and spectral resolution.
   A comparison with the \acp{DIT} currently offered with GRA4MAT shows that we are background limited in low and medium, partly in high, but not in high+.
   In high and high+, the \acp{DIT} could be further increased to improve the S/N.
   However, constraints like dispersion-induced fringe jumps or the need to collect a few DITs per standard \SI{1}{\minute} exposures, favour shorter \acp{DIT}.}
\end{table}

\begin{figure}
   \centering
   \includegraphics[width=\linewidth]{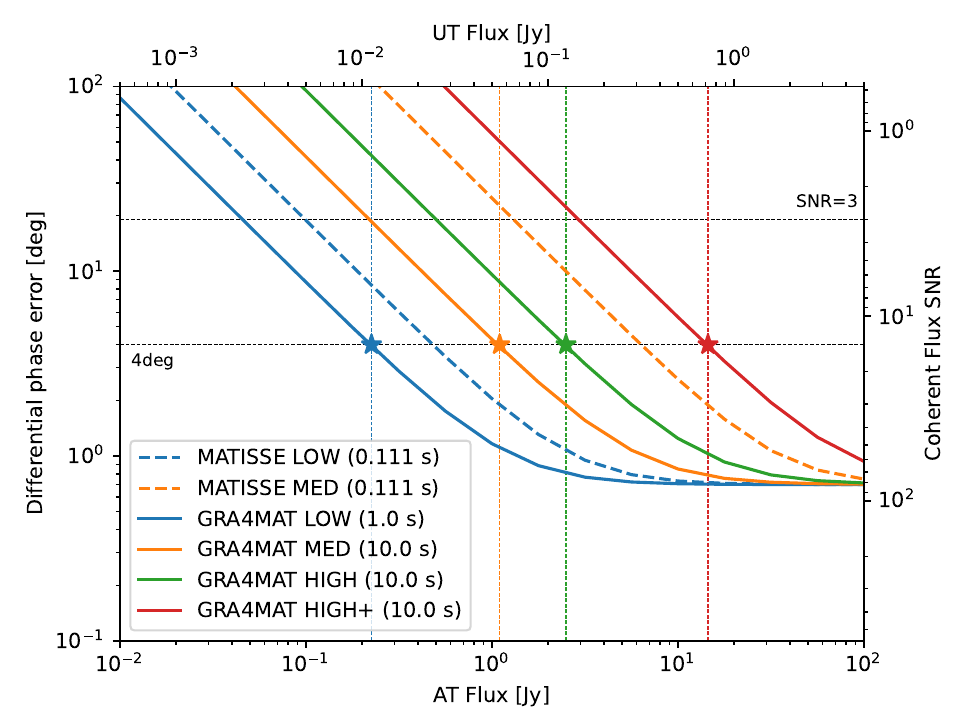}
   \caption{L-band MATISSE (dashed lines) and GRA4MAT (solid lines) differential phase precision (left axis), and associated coherent flux S/N (right axis), per spectral channel, for \SI{1}{\minute} total integration, as a function of the target flux, when observed with the \acp{AT} (bottom axis) or the \acp{UT} (top axis). The coloured lines correspond to the observation-adjusted Exposure Time Calculator predictions for different \ac{DIT} values and spectral resolutions \citep{Petrov+2020}. The highest MATISSE to GRA4MAT $\time10$ gain is observed in medium resolution, where the \ac{DIT} increases transitions from a read noise to a thermal background regime. The high and high+ resolutions are newly offered with GRA4MAT.}
   \label{Fig:MatisseEtcDiffPhase}
\end{figure}

In Fig. \ref{Fig:MatisseEtcDiffPhase} we show a model of the error on the chromatic differential phase in L band as a function of the target flux, for each spectral resolution.
This model is identical to the one used in \citet[][see Fig. 11 therein]{Lopez+2022}, but with \acp{DIT} extended to the GRA4MAT regime.
Its validity was verified against on-sky GRA4MAT observations.
In low spectral resolution, an increase in the detector exposure time from \SI{0.111}{\second} to \SI{1}{\second}, where the background noise dominates, yields a two-fold sensitivity limit improvement.
Following the ESO Call for Proposal convention adopted for MATISSE that considers chromatic phase errors of \SI{4}{\degree} per spectral channel in \SI{1}{\minute} exposure, we reach \SI{\sim0.2}{\Jansky} with the \acp{AT} and \SI{0.01}{\Jansky} with the \acp{UT}.
In medium spectral resolution, the detector exposure time can be increased further, to a background-dominated regime of \SI{10}{\second}, corresponding to a six-fold sensitivity limit improvement.
At higher spectral resolution we remain with a detector exposure time of \SI{10}{\second}, which is around the optimum in high, but well below in high+.
Similar levels of improvement are obtained for the L-band closure phase and for the M-band differential phase and closure phase, but not illustrated here.
We recall that the instrument is perfectly capable of observing targets fainter than the flux levels given above, when the intended science does not require the \SI{4}{\degree} or $\mathrm{S/N}\sim14$ high precision.
At $\mathrm{S/N}=3$ or \SI{\sim20}{\degree}, GRA4MAT reaches \SI{2}{\milli\Jansky} in low spectral resolution with the \acp{UT}, as illustrated in Fig. \ref{Fig:MatisseEtcDiffPhase}.

As already known since \citet{Lopez+2022}, at increasing flux levels the errors on the MATISSE observables decrease and reach a minimum plateau.
The behaviour remains unchanged when MATISSE operates with a fringe tracker, as shown in Fig. \ref{Fig:MatisseEtcDiffPhase}.
This observation does not help us to understand where the effects come from, but it probably rules out the role of residual piston.
This plateau is not investigated further in this work.

\subsection{Reduced N-band coherent flux bias}
\label{SSec:ImprovedBiasN}

\begin{figure}
   \centering
   \includegraphics[width=\linewidth]{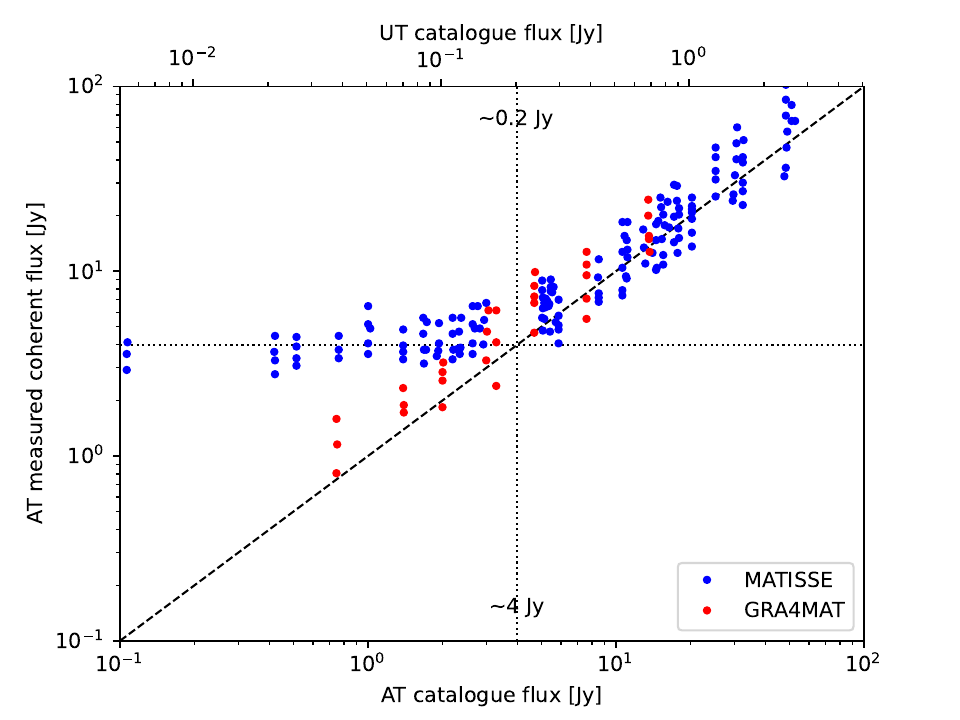}
   \caption{Correlated flux measured with the \acp{AT}, and averaged between \SI{8}{\micro\meter} and \SI{9}{\micro\meter}, as a function of the source flux taken from the MDFC catalogue \citep{Cruzalebes+2019}, without fringe tracker (blue) and with fringe tracker (red).
   The bias generated by the coherent self-integration disappears when fringes are stabilised and the N-band coherent integration is driven by the phase estimated by the fringe tracker.}
   \label{Fig:MatisseBiasBandN}
\end{figure}

Due to the very short exposures and the high thermal background noise, the MATISSE N-band fringes must be coherently integrated in post-processing.
When MATISSE operates independently, this coherent integration is implemented in the pipeline \citep[][section 3.3]{Lopez+2022} and relies on an estimation of the phase from the same data being coherently integrated.
This operation generates a bias at low flux levels, as illustrated in Fig. \ref{Fig:MatisseBiasBandN}.
When operating with the fringe tracker, the fringe phase is measured from the higher S/N K-band data, extrapolated to the N band following the method described in Sect. \ref{SSec:DataReduction}.
In these conditions, the bias is eliminated.
The use of GRA4MAT is therefore always recommended below \SI{\sim5}{\Jansky} with the \acp{AT} and below \SI{\sim300}{\milli\Jansky} with the \acp{UT}, as it allows the respective fundamental (photon, thermal, readout) noise limits of \SI{\sim1}{\Jansky} and \SI{\sim50}{\milli\Jansky} to be reached.

\subsection{Stabilised LM-band transfer function}
\label{SSec:ImprovedTF}

\begin{figure}
   \centering
   \includegraphics[width=\linewidth]{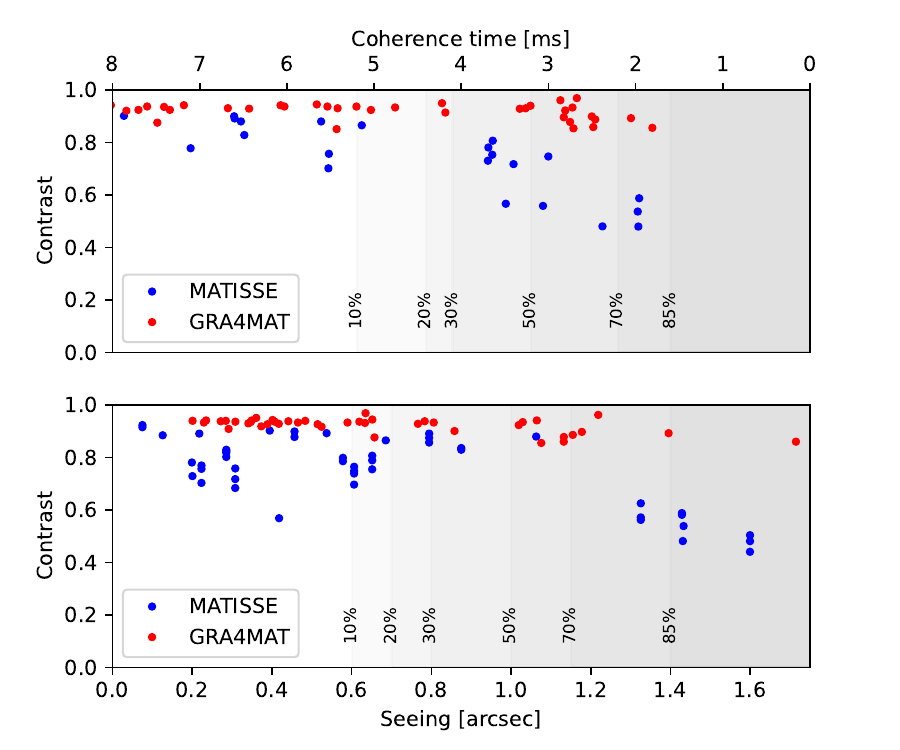}
   \caption{L-band instrumental visibility measured with MATISSE standalone and GRA4MAT on the \acp{AT} vs decreasing coherence time (top) and increasing seeing (bottom). The joint seeing and coherence time probabilities for Paranal are also shown as percentages. The GRA4MAT contrast is higher and more stable in good conditions, and does not degrade as fast in worsening conditions.}
   \label{Fig:MatisseStability}
\end{figure}

The use of GRA4MAT makes the MATISSE LM-band transfer function much less sensitive to seeing and coherence time conditions, as illustrated in Fig. \ref{Fig:MatisseStability}.
The fringe tracker maintains the LM-band instrumental contrast at a level of 0.9 down to a very low coherence time of \SI{\sim2}{\milli\second}, whereas the contrast of MATISSE standalone starts decreasing below a coherence time value of about \SI{4}{\milli\second}.
Seeing values above about \SI{1.2}{\arcsec} show a decrease in the transfer function stability in MATISSE standalone and with GRA4MAT but more marginally.

Figure \ref{Fig:MatisseStability} should not be interpreted as a demonstration that GRA4MAT can be used in poor seeing conditions, however.
The contrasts shown in the figure were selected for MATISSE \acp{DIT} without fringe jumps.
We recall (see section \ref{Sec:FringeTracking}) that the fringe tracker performance depends on the delivery of a stable Strehl by the adaptive optics.

From the dataset presented in Fig.~\ref{Fig:MatisseStability}, we can estimate the raw visibility accuracy as the variability of the contrast in changing atmospheric conditions.
For the 50\% average atmospheric conditions, seeing < \SI{1}{\arcsec} and coherence time longer than \SI{3.2}{\milli\second}, the standard deviation of the contrast is estimated at ±0.08 for MATISSE, and ±0.02 for GRA4MAT.
The MATISSE value matches the estimation given in the first light paper \citep[][Table 5]{Lopez+2022}, whereas the GRA4MAT value is better than the MATISSE standalone performance in good atmospheric conditions.
Additional information on the raw contrast performance is given in Table \ref{Tab:PerformanceSummary}.

\subsection{Visibility measurements}

Absolute visibility measurements with MATISSE require accurate photometry, and hence chopping at frequencies higher than \SI{\sim0.5}{\hertz} for targets fainter than \SI{4}{\Jansky} at \SI{3}{\micro\meter} and \SI{25}{\Jansky} at 4 µm on \acp{AT} (respectively \SI{0.25}{\Jansky} and \SI{1.5}{\Jansky} with \acp{UT}).
This cannot be improved by GRA4MAT.
Only the \acp{DIT} for the low spectral resolution of MATISSE are compatible with this chopping frequency.
However, absolute visibility measurements in the continuum are mandatory only in the same low spectral resolution.
At higher spectral resolutions, MATISSE users are mostly looking for differential measures between spectral features that are not affected by broad-band photometric errors.
Hence, absolute visibility measurements with GRA4MAT should be restricted to low resolution, while higher resolutions with long \acp{DIT} and without chopping should be used only for coherence flux and differential measurements.
Simultaneous chopping and fringe tracking were developed for GRA4MAT in this context (see section \ref{SSec:VisObs}).

\section{Outlook}

The performance of GRAVITY for MATISSE presented here corresponds to what was achieved at the end of the project in mid-2023, and justify the recommendation to always use GRA4MAT over MATISSE standalone whenever possible.
In this last section, we describe the first science observations carried out with the narrow off-axis mode, as it illustrates nicely the newly acquired capabilities of MATISSE.
We then describe how upgrades to the \ac{VLTI}\ac{UT} infrastructure are going to further improve GRA4MAT within the next two years.
Finally, we suggest additional improvements that are currently not in the \ac{VLTI} plan but that would further increase the science relevance of MATISSE.

\subsection{Science highlight: A first exoplanet}

Soon after the first exoplanet discovery \citep{Mayor&Queloz1995}, their observation by differential phase interferometry\footnote{Other interferometric techniques such as indirect detection by wide angle astrometry \citep{Shao&Colavita1992,Colavita+1998,Colavita+1999,Delplancke2008,Woillez+2010} and direct detection by nulling \citep{Serabyn+2012,Absil+2006} have also been proposed and explored, but were not developed here.} was considered \citep{Akeson+1999,Petrov+2001}.
Detection attempts based on the differential phase technique were carried out at VLTI, with the near-infrared AMBER instrument \citep{Millour+2006,Vannier+2006}.
In the mid-infrared specifically, a first attempt was made with simultaneous MIDI and AMBER observations \citep{Matter+2010}.
The goals were to evaluate and calibrate the chromatic effects of air, strongly impacting the mid-infrared domain, and to prepare the approach for future MATISSE observations in this field.
In addition, it was realised at that time that parasitic light effects would introduce biases affecting the planet signal measurement \citep{Matter+2009}; MATISSE was designed to prevent such parasitic effects.
The scientific requirements and conceptual study of MATISSE were taking into account the exigent needs imposed by high-contrast measurements on exoplanets \citep{Lopez+2014,Wolf+2016}. 

The first exoplanet detection by interferometry \citep{GRAVITY+2019} relied on a completely different approach.
The GRAVITY instrument \citep{GRAVITY+2017}, making use of the \SI{2}{\arcsec} delivered by VLTI, positions its single-mode fibres on a slightly off-axis planet for a first coronographic effect on the host star, and uses the angular resolution of the interferometer to further isolate the coherent flux of the planet from the star.
This yields a star-to-planet astrometric position and contrast spectrum.
Drawing inspiration from this GRAVITY success, the first results of exoplanet observations in the mid-infrared at the VLTI were successfully achieved with the narrow off-axis mode of GRA4MAT.

The high sensitivity potential of the off-axis fringe tracking with GRA4MAT was demonstrated on the planet \object{$\beta$ Pictoris b} with MATISSE during the commissioning of the GRA4MAT narrow off-axis mode, on the nights of November 8, 2022, and February 3, 2023 (Prog. ID 60.A-9257).
These observations were motivated by an earlier observing Guaranteed Observing Time proposal (`Retrieving the \object{$\beta$ Pictoris b} spectrum in L band', Prog. ID 110.23SJ), which was submitted to the ESO Observing Program Committee in September 2022, but was not accepted because the mode had not yet been implemented.

\begin{figure*}
   \centering
   \includegraphics[width=\linewidth]{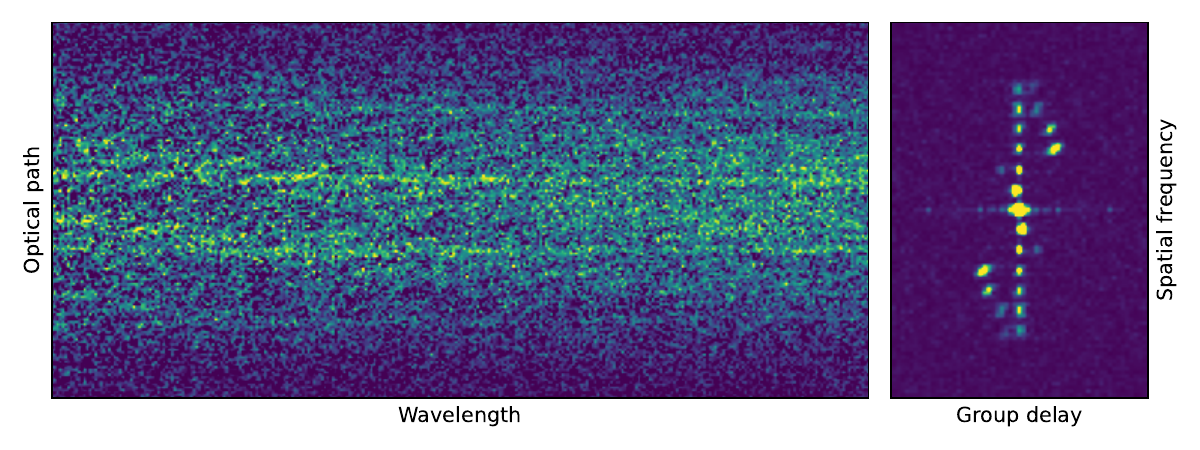}
   \includegraphics[width=\linewidth, trim=0.7cm 0.0cm 0.3cm 0.0cm]{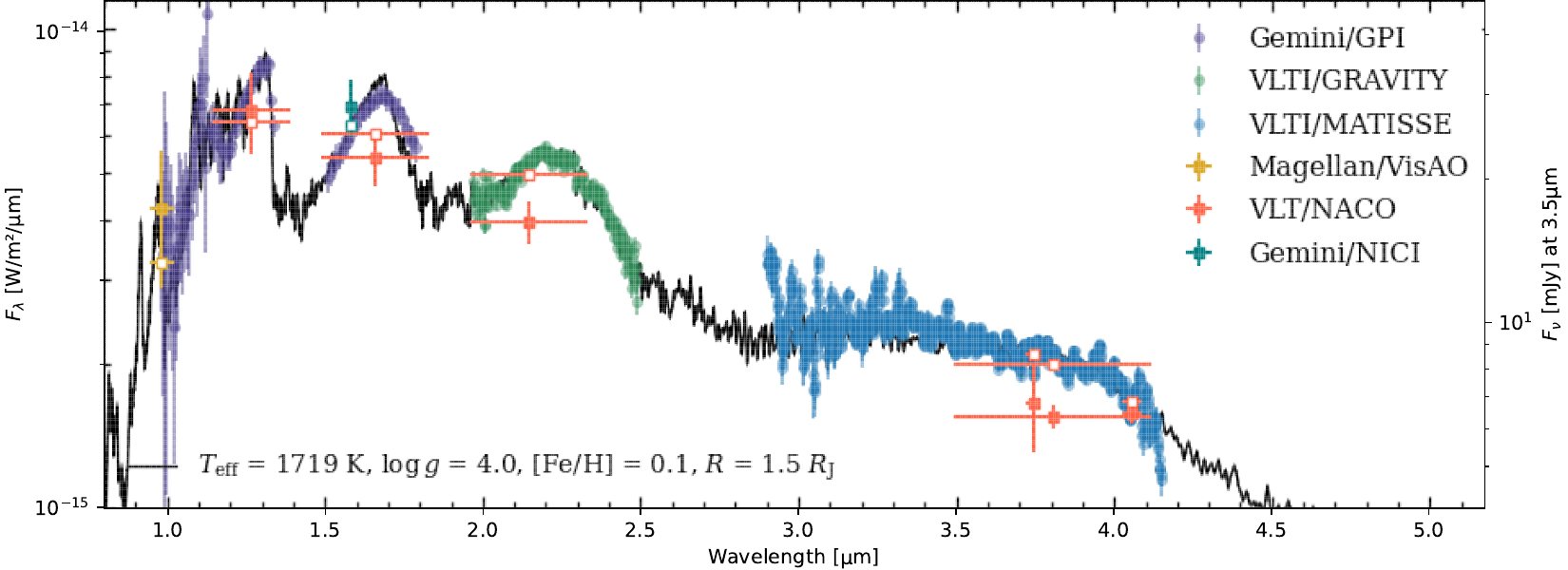}
   \caption{First observation of \object{$\beta$ Pictoris b} with the narrow off-axis mode of GRA4MAT.
            \textbf{Top Left}: Superposition of the exoplanet fringes and host star fringes leaking through the MATISSE pinhole.
            The six exoplanet fringe peaks are vertically aligned at zero group delay.
            The six host star fringe peaks are at various group delay offsets, depending on the observation geometry.
            \textbf{Top right}: 2D power spectral density of the fringes shown in top left panel.
            The central peak represents the total photometry.
            The six vertically aligned peaks correspond to the planet fringes at zero group delay.
            The other six peaks correspond to the star at non-zero group delay.
            \textbf{Bottom}: Preliminary \object{$\beta$ Pictoris b} spectrum (blue), compared to a model Drift-Phenix model (black) and other adaptive optics and interferometric measurements.}
   \label{Fig:BetaPictorisB}
\end{figure*}

We selected the host star \object{$\beta$ Pictoris} a as the fringe-tracking guide star, and alternatively observed the host star and the planet, at the offset predicted by \verb|whereistheplanet| \citep{Wang+2021} which relies on \verb|orbitize!| \citep{Blunt+2019} and the astrometric measurements of \citet{Lacour+2021}.
Figure \ref{Fig:BetaPictorisB} shows the fringes obtained on the planet position where the contributions of the L=\SI{8}{\milli\Jansky} planet and of the star are nearly equivalent.
This corresponds to a 1000-fold attenuation factor for this L=\SI{10.6}{\Jansky} host star at \SI{\sim540}{\mas} ($\sim6\lambda/D$ at \SI{3.5}{\micro\meter}).
A preliminary data reduction derived from the K-band GRAVITY approach of \citet{GRAVITY+2020} yielded a first estimate of the planet spectrum.
To check consistency with a detection of \object{$\beta$ Pictoris b}, we used \verb|species| \citep{Stolker+2020} to fit a grid of Drift-Phenix atmospheric models to a set of photometric and spectroscopic data available in the literature.
We then compared our preliminary L-band spectrum to this best fit, extrapolated over the L band.
We find that the two are in good agreement, as illustrated in Fig.~\ref{Fig:BetaPictorisB}.
A more detailed analysis and interpretation of this observation will be presented in Houllé et al. (in preparation).

\subsection{Planned improvements}
\label{SSec:PlannedImprovements}

The performance of GRA4MAT will continue to improve due to imminent changes to the VLTI infrastructure for the \acp{UT}.
The new \ac{GPAO} for the \ac{VLTI}/\ac{UT}, developed by the GRAVITY+ project \citep{GRAVITY++2022B}, will significantly enhance the behaviour of GRA4MAT on the large telescopes.
Expected in 2024, the \ac{NGS} mode will deliver significantly higher Strehl on brighter objects and make the instrument much less sensitive to degraded seeing conditions.
We expect the fringe tracker operation to become fringe-jump-free on the \acp{UT}, just as it currently is on the \acp{AT}.
When the \ac{LGS} mode of \ac{GPAO} becomes available, good Strehl conditions will be obtained on targets that are much fainter in the visible.
The fringe tracking limiting magnitude mismatch between \acp{AT} and \acp{UT} will then be remedied.
At a foreseen limiting magnitude of K=\SI{12.5}{\mag}, and virtually no limit on the \ac{AO} tip-tilt star at GAIA G=\SI{17}{\mag}, GRA4MAT will be able to systematically observe red and faint targets, such as \acp{YSO} and \acp{AGN}.

We also expect that the higher and more stable Strehl delivered to the fringe tracker will make GRA4MAT compatible with chopping on the \acp{UT}.
This capability will enable the high-precision visibility measurements, in medium resolution and above, required for sensitive mid-infrared interferometric imaging with the larger telescopes.

As explained in section \ref{SSec:AcquisitionSequence}, GRA4MAT uses the main VLTI delay lines as common-path piston actuators, and is affected by from a \SI{50}{\micro\second} latency penalty (see Fig. \ref{Fig:Latency}).
This penalty will soon be removed by using the \acp{BCDDL} recently installed by the GRAVITY+ project.
These additional common-path delay lines, primarily developed as improved \acp{DDL} for GRAVITY wide off-axis observations \citep{GRAVITY++2022A}, provide the same level of performance as the \ac{TTP} internal actuators of the GRAVITY fringe tracker.

\subsection{Possible future developments}

At the time of writing this paper, with the exception of section \ref{SSec:PlannedImprovements}, the GRA4MAT project at ESO was concluded.
What follows are possible developments that should be considered.

The optical path control strategy could be further improved by including a continuous dispersion tracking capability in the fringe tracker.
This would marginally improve the fringe contrast on longer \acp{DIT} and would also reduce fringe jumps observed on nights with high water vapour content (column larger than \SI{\sim5}{\milli\meter}).

The GRA4MAT implementation could benefit from an additional fast piston actuator not in the common path, but exclusive for MATISSE.
With such a set-up, the GRAVITY fringe tracker could send an open-loop feed-forward command, improving further the OPD rejection at low frequencies.
This control strategy was prototyped \citep{Lane&Colavita2003} on the Palomar Testbed Interferometer \citep{Colavita+1999} and operated with the Keck Nuller \citep{Colavita+2010} on the Keck Interferometer \citep{Colavita+2013}.
The benefit for MATISSE, a higher and more stable fringe contrast, might not be as strong as for a high-precision nulling instrument, for example NOTT \citep{Defrere+2018}.

GRA4MAT would also benefit from further developing the MATISSE pipeline.
Coherent integrations in post-processing over durations well beyond the \acp{DIT} are already possible in the N band.
In the LM bands, however, longer than \SI{10}{\second} coherent integrations must be implemented, especially for observations involving the more sensitive \acp{UT}.
The \object{$\beta$ Pictoris b} results shown in Fig. \ref{Fig:BetaPictorisB} actually demonstrate that integrations beyond the \acp{DIT} are possible.
Based on Fig. \ref{Fig:MatisseEtcDiffPhase}, the instantaneous coherent flux S/N for this \SI{8}{\milli\Jansky} target is $\sim2.5$, whereas the contrast spectrum derived from the combination of many \acp{DIT} reaches a S/N of $\sim80$.
We hope that one day MATISSE becomes capable of measurements on targets virtually invisible in individual LM-band \acp{DIT}.

We believe that adding a wide off-axis capability to MATISSE has a strong scientific potential.
A MATISSE wide implementation would not require any new hardware developments.
The dual-field building blocks are already present at VLTI and actively used by GRAVITY wide \citep{GRAVITY++2022A}.
MATISSE and the infrared tip-tilt sensor IRIS can already receive the off-axis beams while keeping GRAVITY on axis.
MATISSE wide would just be an extension of the GRA4MAT control software and the commissioning of a new mode.
Considering that the atmospheric turbulence retains coherence at larger off-axis distances in the mid-infrared, MATISSE wide would drastically expand the MATISSE sky coverage.
We would then be able to start observation campaigns of emblematic AGN and YSO science cases \citep[see e.g.][]{Boskri+2021}.

\section{Conclusions}

\begin{table*}
   \centering
   \caption{Performance and capabilities improved from MATISSE to current GRA4MAT, and future planned improvements when \ac{GPAO} becomes available on the \acp{UT}.}
   \label{Tab:PerformanceSummary}
   \fontsize{8}{9.5}\selectfont
   \begin{tabular}{|c|c|cc|cc|}
      \hline
      Metric & MATISSE & \multicolumn{2}{|c|}{GRA4MAT} & \multicolumn{2}{|c|}{GRA4MAT+GPAO} \\
      \hline
      \hline
      \rowcolor{LightGray}
      \multicolumn{6}{|c|}{L-band spectral resolution and instantaneous coverage} \\
      \hline
      LM low (R$\sim$35, \SIrange{2.9}{5.0}{\micro\meter})     & 100\%                         & 100\% & \multirow{4}{*}{\S\ref{SSec:ImprovedCoverage}} && \\
      LM medium (R$\sim$500, \SIrange{2.9}{5.0}{\micro\meter}) & 6.4\% \SI{0.16}{\micro\meter} & 100\% &&& \\
      L high (R$\sim$1000, \SIrange{2.9}{4.2}{\micro\meter})   & 6.4\% \SI{0.8}{\micro\meter}  & 100\% &&& \\
      L high+ (R$\sim$3300, \SIrange{4.02}{5.0}{\micro\meter}) &  0\%                          & 100\% &&& \\
      \hline
      \hline
      \rowcolor{LightGray}
      \multicolumn{6}{|c|}{Detector integration time [\si{\second}]} \\
      \hline
      L low    &      1      & 1   & \multirow{4}{*}{\S\ref{SSec:ImprovedLMSensitivity}} && \\
      L medium &     8.5     & 10  &&& \\
      L high   & not offered & 10  &&& \\
      L high+  & not offered & 10  &&& \\
      \hline
      M low    &      2      & 0.7 & \multirow{3}{*}{\S\ref{SSec:ImprovedLMSensitivity}} && \\
      M medium &     13      & 11  &&& \\
      M high+  & not offered & 17  &&& \\
      \hline
      N low    &     17      & 3   &&& \\
      N high   & not offered & 26  &&& \\
      \hline
      \hline
      \rowcolor{LightGray}
      \multicolumn{6}{|c|}{AT correlated flux sensitivity within 30\% best conditions [\si{\Jansky}]} \\
      \hline
      L low    &      1      & 0.2 & \multirow{4}{*}{\S\ref{SSec:ImprovedLMSensitivity}} && \\
      L medium &     8.5     & 1   &&& \\
      L high   & not offered & 2   &&& \\
      L high+  & not offered & 10  &&& \\
      \hline
      M low    &      2      & 0.7 & \multirow{3}{*}{\S\ref{SSec:ImprovedLMSensitivity}} && \\
      M medium &     13      & 11  &&& \\
      M high+  & not offered & 17  &&& \\
      \hline
      N low    &     17      & 3   & \multirow{2}{*}{\S\ref{SSec:ImprovedBiasN}}  && \\
      N high   & not offered & 26  &&& \\
      \hline
      \hline
      \rowcolor{LightGray}
      \multicolumn{6}{|c|}{UT correlated flux sensitivity within 30\% best conditions [\si{\Jansky}]} \\
      \hline
      L low    &    0.06     & 0.02 & \multirow{4}{*}{\S\ref{SSec:ImprovedLMSensitivity}} & 0.01 & \multirow{4}{*}{\S\ref{SSec:PlannedImprovements}} \\
      L medium &    0.6      & 0.3  && 0.05 & \\
      L high   & not offered & 0.7  && 0.1  & \\
      L high+  & not offered & 6    && 0.5  & \\
      \hline
      M low    &    0.5      & 0.07 & \multirow{3}{*}{\S\ref{SSec:ImprovedLMSensitivity}} & 0.035 & \multirow{3}{*}{\S\ref{SSec:PlannedImprovements}} \\
      M medium &    1.3      & 1    && 0.55 & \\
      M high+  & not offered & 6    && 0.85 & \\
      \hline
      N low    &    0.9      & 0.1  & \multirow{2}{*}{\S\ref{SSec:ImprovedBiasN}} && \\
      N high   & not offered & 0.5  &&& \\
      \hline
      \hline
      \rowcolor{LightGray}
      \multicolumn{6}{|c|}{Raw contrast} \\
      \hline
      \hline
      30\% best conditions & $0.85\pm0.06$ & $0.93\pm0.02$ & \multirow{2}{*}{\S\ref{SSec:ImprovedTF}} && \\
      50\% best conditions & $0.81\pm0.08$ & $0.93\pm0.02$ &&& \\
      \hline
      \rowcolor{LightGray}
      \multicolumn{6}{|c|}{Observing capabilities} \\
      \hline
      AT chopping: photometry \& visibility & Yes & Yes &&&     \\
      UT chopping: photometry \& visibility & Yes & No & \S\ref{SSec:VisObs} & Yes & \S\ref{SSec:PlannedImprovements} \\
      Narrow off-axis observations & No  & Yes & \S\ref{Sec:NarrowOffAxisOffsets} &&    \\
      \hline
   \end{tabular}
   \tablefoot{Detailed explanations are available in the referenced \S{X.X} sections. More updated values, covering a wider range of atmospheric conditions, are available on the VLTI/MATISSE instrument page of the ESO website (\url{https://www.eso.org/sci/facilities/paranal/instruments/matisse.html}).}
\end{table*}

The GRAVITY fringe tracker is now used by both the GRAVITY and MATISSE instruments.
As a result, MATISSE saved costs and benefited from the fringe tracking expertise developed by the GRAVITY team.
Likewise, GRAVITY benefited from a single focus on its fringe tracker, avoiding dispersion and effort duplication.
For example, MATISSE motivated the fringe jump control feature, which is also an added value for GRAVITY for high-precision differential visibility across the K band.
Therefore, selecting one shared fringe tracker federated the teams around one single project, and made it better for all.
This illustrates the value of a mutualisation strategy.

Thanks to GRA4MAT, MATISSE now delivers its predicted performance.
A summary is given in Table~\ref{Tab:PerformanceSummary} of all the performance improvements, and of the performance improvements expected with the \acp{UT}, when \ac{GPAO} becomes available in late 2024.
A second wave of astrophysical MATISSE observations involving sensitivity, spectroscopy, accuracy, and contrast are now in progress; the associated publications are on their way.
The first narrow off-axis observations with MATISSE are opening a new field of mid-infrared extrasolar planet observations with VLTI \citep{Houlle+2024}.

For the past 20 years, single-telescope instrumentation for high angular resolution and high-contrast observations has been unimaginable without the support of an \ac{AO} system \citep{Wizinowich+2000,Brandner+2002}.
A similar revolution is now happening for optical long-baseline interferometry.
This paradigm shift reveals itself clearly when comparing early interferometry reviews \citep{Quirrenbach+2001,Monnier2003} to more recent ones \citep{Eisenhauer+2023}.
The interferometric instruments, that used to focus on bright targets within the tight coherence limits of the atmosphere, cannot be conceived any more without fringe tracking support.
Initiated by the GRAVITY instrument \citep{GRAVITY+2017}, GRAVITY for MATISSE has now completed this revolution for all modern VLTI instruments.

\begin{acknowledgements}
   Based on observations collected at the European Southern Observatory under ESO commissioning programme 60.A-9257.
   This research has made use of the Jean-Marie Mariotti Center \texttt{Aspro} service.\footnote{Available at http://www.jmmc.fr/aspro}
   P. Berio, M. Houllé, S. Lacour and F. Millour acknowledge the support of the French Agence Nationale de la Recherche (ANR), under grant ANR-21-CE31-0017 (project ExoVLTI).
   J. Leftley and R. Petrov acknowledge the support of the french ANR under the grant AGN\_MELBa ANR ANR-21-CE31-001.
   A. Meilland is supported by the ANR grant MASSIF (ANR-21-CE31-0018).
   J. Varga is funded from the Hungarian NKFIH OTKA project no. K-132406. J. Varga has received funding from the European Research Council (ERC) under the European Union's Horizon 2020 research and innovation programme under grant agreement No 716155 (SACCRED). The research of J. Varga is supported by NOVA, the Netherlands Research School for Astronomy. J. Varga acknowledges support from the Fizeau exchange visitors program. The research leading to these results has received funding from the European Union's Horizon 2020 research and innovation programme under Grant Agreement 101004719 (ORP).
\end{acknowledgements}

\bibliographystyle{aa}
\bibliography{gra4mat}

\begin{appendix}
\end{appendix}

\end{document}